\documentclass[aps,prl,superscriptaddress,showpacs,preprintnumbers,amsmath,amssymb]{revtex4-1}
\def\Kst{K^*(892)}
\def\Mbc{M_{\rm bc}}

\usepackage{graphicx} 
\usepackage{dcolumn}  
\usepackage[left]{lineno}
\PassOptionsToPackage{hyphens}{url}
\usepackage[colorlinks]{hyperref}
\usepackage[hyphenbreaks]{breakurl}
\graphicspath{{ps}}

\begin{document}



\title{ \quad\\[1.0cm] Evidence for Isospin Violation and Measurement of $CP$ Asymmetries in $B \to \Kst \gamma$}

\noaffiliation
\affiliation{University of the Basque Country UPV/EHU, 48080 Bilbao}
\affiliation{Beihang University, Beijing 100191}
\affiliation{Budker Institute of Nuclear Physics SB RAS, Novosibirsk 630090}
\affiliation{Faculty of Mathematics and Physics, Charles University, 121 16 Prague}
\affiliation{Chonnam National University, Kwangju 660-701}
\affiliation{University of Cincinnati, Cincinnati, Ohio 45221}
\affiliation{Deutsches Elektronen--Synchrotron, 22607 Hamburg}
\affiliation{University of Florida, Gainesville, Florida 32611}
\affiliation{Justus-Liebig-Universit\"at Gie\ss{}en, 35392 Gie\ss{}en}
\affiliation{SOKENDAI (The Graduate University for Advanced Studies), Hayama 240-0193}
\affiliation{Hanyang University, Seoul 133-791}
\affiliation{University of Hawaii, Honolulu, Hawaii 96822}
\affiliation{High Energy Accelerator Research Organization (KEK), Tsukuba 305-0801}
\affiliation{J-PARC Branch, KEK Theory Center, High Energy Accelerator Research Organization (KEK), Tsukuba 305-0801}
\affiliation{IKERBASQUE, Basque Foundation for Science, 48013 Bilbao}
\affiliation{Indian Institute of Science Education and Research Mohali, SAS Nagar, 140306}
\affiliation{Indian Institute of Technology Bhubaneswar, Satya Nagar 751007}
\affiliation{Indian Institute of Technology Guwahati, Assam 781039}
\affiliation{Indian Institute of Technology Madras, Chennai 600036}
\affiliation{Indiana University, Bloomington, Indiana 47408}
\affiliation{Institute of High Energy Physics, Chinese Academy of Sciences, Beijing 100049}
\affiliation{Institute of High Energy Physics, Vienna 1050}
\affiliation{Institute for High Energy Physics, Protvino 142281}
\affiliation{INFN - Sezione di Napoli, 80126 Napoli}
\affiliation{INFN - Sezione di Torino, 10125 Torino}
\affiliation{Advanced Science Research Center, Japan Atomic Energy Agency, Naka 319-1195}
\affiliation{J. Stefan Institute, 1000 Ljubljana}
\affiliation{Kanagawa University, Yokohama 221-8686}
\affiliation{Institut f\"ur Experimentelle Kernphysik, Karlsruher Institut f\"ur Technologie, 76131 Karlsruhe}
\affiliation{Kavli Institute for the Physics and Mathematics of the Universe (WPI), University of Tokyo, Kashiwa 277-8583}
\affiliation{Kennesaw State University, Kennesaw, Georgia 30144}
\affiliation{King Abdulaziz City for Science and Technology, Riyadh 11442}
\affiliation{Department of Physics, Faculty of Science, King Abdulaziz University, Jeddah 21589}
\affiliation{Korea Institute of Science and Technology Information, Daejeon 305-806}
\affiliation{Korea University, Seoul 136-713}
\affiliation{Kyungpook National University, Daegu 702-701}
\affiliation{\'Ecole Polytechnique F\'ed\'erale de Lausanne (EPFL), Lausanne 1015}
\affiliation{P.N. Lebedev Physical Institute of the Russian Academy of Sciences, Moscow 119991}
\affiliation{Faculty of Mathematics and Physics, University of Ljubljana, 1000 Ljubljana}
\affiliation{Ludwig Maximilians University, 80539 Munich}
\affiliation{Luther College, Decorah, Iowa 52101}
\affiliation{University of Maribor, 2000 Maribor}
\affiliation{Max-Planck-Institut f\"ur Physik, 80805 M\"unchen}
\affiliation{School of Physics, University of Melbourne, Victoria 3010}
\affiliation{University of Miyazaki, Miyazaki 889-2192}
\affiliation{Moscow Physical Engineering Institute, Moscow 115409}
\affiliation{Moscow Institute of Physics and Technology, Moscow Region 141700}
\affiliation{Graduate School of Science, Nagoya University, Nagoya 464-8602}
\affiliation{Kobayashi-Maskawa Institute, Nagoya University, Nagoya 464-8602}
\affiliation{Nara Women's University, Nara 630-8506}
\affiliation{National Central University, Chung-li 32054}
\affiliation{National United University, Miao Li 36003}
\affiliation{Department of Physics, National Taiwan University, Taipei 10617}
\affiliation{H. Niewodniczanski Institute of Nuclear Physics, Krakow 31-342}
\affiliation{Nippon Dental University, Niigata 951-8580}
\affiliation{Niigata University, Niigata 950-2181}
\affiliation{Novosibirsk State University, Novosibirsk 630090}
\affiliation{Osaka City University, Osaka 558-8585}
\affiliation{Pacific Northwest National Laboratory, Richland, Washington 99352}
\affiliation{University of Pittsburgh, Pittsburgh, Pennsylvania 15260}
\affiliation{Punjab Agricultural University, Ludhiana 141004}
\affiliation{Theoretical Research Division, Nishina Center, RIKEN, Saitama 351-0198}
\affiliation{University of Science and Technology of China, Hefei 230026}
\affiliation{Showa Pharmaceutical University, Tokyo 194-8543}
\affiliation{Soongsil University, Seoul 156-743}
\affiliation{Sungkyunkwan University, Suwon 440-746}
\affiliation{School of Physics, University of Sydney, New South Wales 2006}
\affiliation{Department of Physics, Faculty of Science, University of Tabuk, Tabuk 71451}
\affiliation{Tata Institute of Fundamental Research, Mumbai 400005}
\affiliation{Excellence Cluster Universe, Technische Universit\"at M\"unchen, 85748 Garching}
\affiliation{Department of Physics, Technische Universit\"at M\"unchen, 85748 Garching}
\affiliation{Toho University, Funabashi 274-8510}
\affiliation{Department of Physics, Tohoku University, Sendai 980-8578}
\affiliation{Earthquake Research Institute, University of Tokyo, Tokyo 113-0032}
\affiliation{Department of Physics, University of Tokyo, Tokyo 113-0033}
\affiliation{Tokyo Institute of Technology, Tokyo 152-8550}
\affiliation{Tokyo Metropolitan University, Tokyo 192-0397}
\affiliation{University of Torino, 10124 Torino}
\affiliation{Utkal University, Bhubaneswar 751004}
\affiliation{Virginia Polytechnic Institute and State University, Blacksburg, Virginia 24061}
\affiliation{Wayne State University, Detroit, Michigan 48202}
\affiliation{Yamagata University, Yamagata 990-8560}
\affiliation{Yonsei University, Seoul 120-749}
  \author{T.~Horiguchi}\affiliation{Department of Physics, Tohoku University, Sendai 980-8578} 
  \author{A.~Ishikawa}\affiliation{Department of Physics, Tohoku University, Sendai 980-8578} 
  \author{H.~Yamamoto}\affiliation{Department of Physics, Tohoku University, Sendai 980-8578} 
  \author{I.~Adachi}\affiliation{High Energy Accelerator Research Organization (KEK), Tsukuba 305-0801}\affiliation{SOKENDAI (The Graduate University for Advanced Studies), Hayama 240-0193} 
  \author{H.~Aihara}\affiliation{Department of Physics, University of Tokyo, Tokyo 113-0033} 
  \author{S.~Al~Said}\affiliation{Department of Physics, Faculty of Science, University of Tabuk, Tabuk 71451}\affiliation{Department of Physics, Faculty of Science, King Abdulaziz University, Jeddah 21589} 
  \author{D.~M.~Asner}\affiliation{Pacific Northwest National Laboratory, Richland, Washington 99352} 
  \author{V.~Aulchenko}\affiliation{Budker Institute of Nuclear Physics SB RAS, Novosibirsk 630090}\affiliation{Novosibirsk State University, Novosibirsk 630090} 
  \author{T.~Aushev}\affiliation{Moscow Institute of Physics and Technology, Moscow Region 141700} 
  \author{R.~Ayad}\affiliation{Department of Physics, Faculty of Science, University of Tabuk, Tabuk 71451} 
% ...
\collaboration{The Belle Collaboration}


\begin{abstract}
We report the first evidence for isospin violation in $B \to K^* \gamma$ and the first measurement of difference of $CP$ asymmetries between $B^+ \to K^{*+} \gamma$ and $B^0 \to K^{*0} \gamma$. This analysis is based on the data sample containing $772 \times 10^6 B\bar{B}$ pairs that was collected with the Belle detector at the KEKB energy-asymmetric $e^+ e^-$ collider. We find evidence for the isospin violation with a significance of 3.1$\sigma$, $\Delta_{0+} = (+6.2 \pm 1.5 ({\rm stat.}) \pm 0.6 ({\rm syst.}) \pm 1.2 (f_{+-}/f_{00}))$\%, where the third uncertainty is due to the uncertainty on the fraction of $B^+B^-$ to $B^0\bar{B}^0$ production in $\Upsilon(4S)$ decays. The measured value is consistent with predictions of the SM. The result for the difference of $CP$ asymmetries is $\Delta A_{CP} = (+2.4 \pm 2.8({\rm stat.}) \pm 0.5({\rm syst.}))$\%, consistent with zero.  The measured branching fractions and $CP$ asymmetries for charged and neutral $B$ meson decays are the most precise to date. We also calculate the ratio of branching fractions of $B^0 \to K^{*0} \gamma$ to $B_s^0 \to \phi \gamma$.
\end{abstract}


\maketitle

\tighten

{\renewcommand{\thefootnote}{\fnsymbol{footnote}}}
\setcounter{footnote}{0}

Radiative $b \to s \gamma$ decays proceed predominantly via one-loop electromagnetic penguin diagrams. This process is also possible via annihilation diagrams; however, the amplitudes are highly suppressed by ${\cal{O}}(\Lambda_{\rm QCD}/m_b)$ and CKM matrix elements~\cite{Cabibbo:1963yz,Kobayashi:1973fv} in the Standard Model~(SM)~\cite{Bosch:2001gv,Beneke:2000wa}. Since new heavy particles could contribute to the loops, the $b \to s \gamma$ process is a sensitive probe for new physics~(NP). Furthermore, new particles could mediate the annihilation diagrams or effective four-fermion contact interactions with different magnitudes in charged and neutral $B$ meson decays, so that the penguin dominance in $b \to s \gamma$ might be violated. The $B \to K^* \gamma$ decay~\cite{Kstar} is experimentally the cleanest exclusive decay mode among the $B \to X_s \gamma$ decays. The branching fractions give weak constraints on NP since the SM predictions suffer from large uncertainties in the form factors, while the isospin~($\Delta_{0+}$) and direct $CP$ asymmetries~($A_{CP}$) are theoretically clean observables due to cancellation of these uncertainties~\cite{Keum:2004is}. The $\Delta_{0+}$, $A_{CP}$, and difference and average of $A_{CP}$ between charged and neutral $B$ mesons ($\Delta A_{CP}$ and $\bar{A}_{CP}$) are defined as 
\begin{eqnarray}
&&\Delta_{0+}   = \frac{\Gamma(B^0 \to K^{*0} \gamma)-\Gamma(B^+ \to K^{*+} \gamma)}{\Gamma(B^0 \to K^{*0} \gamma)+\Gamma(B^+ \to K^{*+} \gamma)},\\
&&A_{CP}        = \frac{\Gamma(\bar{B} \to \bar{K}^* \gamma)-\Gamma(B \to K^* \gamma)}{\Gamma(\bar{B} \to \bar{K}^* \gamma)+\Gamma(B \to K^* \gamma)},\\
&&\Delta A_{CP} = A_{CP}(B^+ \to K^{*+} \gamma) - A_{CP}(B^0 \to K^{*0} \gamma),\\
&&\bar{A}_{CP} = \frac{A_{CP}(B^+ \to K^{*+} \gamma) + A_{CP}(B^0 \to K^{*0} \gamma)}{2},\\
&&\frac{\Gamma(B^0 \to K^{*0} \gamma)}{\Gamma(B^+ \to K^{*+} \gamma)} = \frac{\tau_{B^+}}{\tau_{B^0}}\frac{f_{+-}}{f_{00}} \frac{N(B^0 \to K^{*0} \gamma)}{N(B^+ \to K^{*+} \gamma)},
\end{eqnarray}
where the $\Gamma$ denotes the partial width, $N$ is the number of produced signal events, $\tau_{B^+}/\tau_{B^0}$ is the lifetime ratio of $B^+$ to $B^0$ mesons, and $f_{+-}$ and $f_{00}$ are the $\Upsilon(4S)$ branching fractions to $B^+B^-$ and $B^0\bar{B}^0$ decays, respectively. Predictions of the isospin asymmetry range from 2\% to 8\% with a typical uncertainty of 2\% in the SM~\cite{Lyon:2013gba,Ball:2006eu,Keum:2004is,Kagan:2001zk,Ahmady:2013cva,Jung:2012vu}, while a large deviation from the SM predictions is possible due to NP~\cite{Lyon:2013gba,Kagan:2001zk,Jung:2012vu}. $A_{CP}$ is predicted to be small in the SM~\cite{Keum:2004is,Paul:2016urs,Greub:1994tb,Jung:2012vu}; hence, a measurement of $CP$ violation is a good probe for NP~\cite{Dariescu:2007gr}. The isospin difference of direct $CP$ violation is theoretically discussed in the context of inclusive $B \to X_s \gamma$ process~\cite{Benzke:2010tq} but heretofore not in the exclusive $B \to K^* \gamma$ channel; however, $\Delta A_{CP}$ here will be useful to identify NP once $A_{CP}$ is observed. 

The $B \to K^* \gamma$ decays were studied by CLEO~\cite{Coan:1999kh}, Belle~\cite{Nakao:2004th}, Babar~\cite{Aubert:2009ak} and LHCb~\cite{Aaij:2012ita}. The current world averages of the isospin and direct $CP$ asymmetries are $\Delta_{0+} = (+5.2 \pm 2.6)$\%, $A_{CP}(B^0 \to K^{*0} \gamma) = (-0.2\pm1.5)$\%, $A_{CP}(B^+ \to K^{*+} \gamma) = (+1.8 \pm 2.9)$\% and $A_{CP}(B \to K^{*} \gamma) = (-0.3 \pm 1.7)$\%~\cite{Olive:2016xmw}, respectively, which are consistent with predictions in the SM and give strong constraints on NP~\cite{Paul:2016urs,Altmannshofer:2014rta,DescotesGenon:2011yn,Mahmoudi:2007gd,Jung:2012vu}. The world averages of branching fractions are also consistent with predictions within the SM~\cite{Straub:2015ica,Ali:2007sj,Ali:2001ez,Ball:2006eu,Keum:2004is,Bosch:2001gv,Greub:1994tb,Jung:2012vu} and are used for constraining NP~\cite{Paul:2016urs,Ciuchini:2016weo,Jung:2012vu}.

In this Letter, we report the first evidence of isospin violation in $B \to K^* \gamma$. In addition, we present measurements of the branching fractions, direct $CP$ asymmetries and their isospin difference and average. We use the full $\Upsilon(4S)$ resonance data sample collected by the Belle detector at the KEKB energy-asymmetric collider~\cite{KEKB}; this sample contains $772 \times 10^6 B\bar{B}$ pairs. The results supersede our previous measurements~\cite{Nakao:2004th}.

The Belle detector is a large-solid-angle magnetic
spectrometer that consists of a silicon vertex detector (SVD),
a 50-layer central drift chamber (CDC), an array of
aerogel threshold Cherenkov counters (ACC),  
a barrel-like arrangement of time-of-flight
scintillation counters (TOF), and an electromagnetic calorimeter
comprised of CsI(Tl) crystals (ECL) located inside 
a super-conducting solenoid coil that provides a 1.5~T
magnetic field.  An iron flux-return located outside of
the coil is instrumented to detect $K_L^0$ mesons and to identify
muons. The $z$ axis is aligned with the direction opposite the $e^+$ beam.
The detector is described in detail elsewhere~\cite{Belle}.

The selection is optimized with Monte Carlo~(MC) simulation samples. The MC events are generated with EvtGen~\cite{EvtGen} and the detector simulation is done by GEANT3~\cite{GEANT}. We reconstruct $B^0 \to K^{*0} \gamma$ and $B^+ \to K^{*+} \gamma$ decays, where $K^*$ is formed from $K^+ \pi^-$, $K_S^0 \pi^0$, $K^+ \pi^0$ or $K_S^0 \pi^+$ combinations~\cite{CC}.

Prompt photon candidates are selected from isolated clusters in the ECL that are not associated with any charged tracks reconstructed by the SVD and the CDC. We require the ratio of the energy deposited in a $3\times3$ array of ECL crystals centered on the crystal having the maximum energy to that in the enclosing $5\times5$ array to be above 0.95. 
The photon energy in the center-of-mass~(CM) frame is required to be in the range of 1.8~GeV $< E_{\gamma}^* <$ 3.4~GeV.
The polar angle of the photon candidate is required to be in the barrel region of the ECL ($33^{\circ} < \theta_{\gamma} < 128^{\circ}$) to take advantage of the better energy resolution in the barrel compared with the endcap and to reduce continuum $e^+e^- \to q\bar{q}$ $(q=u,d,s,c)$ background with initial state radiation. 
The dominant backgrounds to the prompt photons are from asymmetric-energy decays of high momentum $\pi^0$ or $\eta$ mesons, where one photon is hard and the other is soft. These events can be suppressed by using two probability density functions~(PDFs) for $\pi^0$ and $\eta$ constructed from the following two variables: the invariant mass of the photon candidate and another photon in an event, and the energy of this additional photon in laboratory frame. We require that the $\pi^0$ and $\eta$ probabilities are less than 0.3. These requirements retain about 92\% of signal events while removing about 61\% of continuum background.

To reject misreconstructed tracks and beam backgrounds, charged tracks except for the $K_S^0 \to \pi^+ \pi^-$ decay daughters are required to have a momentum in the laboratory frame greater than 0.1~GeV/$c$. In addition, we require that the impact parameter with respect to the nominal interaction point~(IP) be less than 0.5~cm transverse to, and 5.0~cm along, the $z$ axis. 
To identify $K^+$ and $\pi^+$, a likelihood ratio is calculated from the specific ionization measurements in the CDC, time-of-flight information from the TOF and the response of the ACC. 

$K_S^0$ candidates are reconstructed from pairs of oppositely-charged tracks, treated as pions, and identified by a
multivariate analysis with a neural network~\cite{NB} based on two sets of input variables~\cite{nakano}. 
The first set of variables, which separate $K_S^0$ candidates from combinatorial background, are:
(1) the $K_S^0$ momentum in the laboratory frame,
(2) the distance along the $z$ axis between the two track helices at their closest approach,
(3) the flight length in the $x$-$y$ plane,
(4) the angle between the $K_S^0$ momentum and the vector joining the $K_S^0$ decay vertex and the nominal IP,
(5) the angle between the $\pi$ momentum and the laboratory-frame direction of the $K_S^0$ in the $K_S^0$ rest frame,
(6) the distance of closest approach in the $x$-$y$ plane between the nominal IP and the pion helices,
and (7) the pion hit information in the SVD and CDC.
The second set of variables, which identify $\Lambda \to p\pi^-$ background, are:
(1) particle identification information, momentum and polar angles of the two daughter tracks,
and (2) invariant mass with the proton- and pion-mass hypotheses.
In addition, the $K^0_S$ candidate is required to have an invariant mass $M_{\pi\pi}$, calculated with the pion-mass hypothesis, that satisfies $|M_{\pi\pi} - m_{K_S^0}| < 10$~MeV/$c^2$, where $m_{K^0_S}$ is the nominal $K^0_S$ mass; this requirement corresponds to a $\pm$3$\sigma$ interval in mass resolution.

We reconstruct $\pi^0$ candidates from two photons each with energy greater than 50~MeV.
We require the invariant mass to be within $\pm10$~MeV/$c^2$ of the nominal $\pi^0$ mass, corresponding to about 2$\sigma$ in resolution. To reduce the large combinatorial background, we require that the $\pi^0$ momentum in the CM frame, calculated with a $\pi^0$ mass-constraint fit, be greater than 0.5~GeV/$c$ and the cosine of the angle between two photons be greater than 0.5.

$K^*$ candidates are selected with a loose invariant mass selection of $M_{K\pi} < 2.0$~GeV/$c^2$.

  $B$ meson candidates are reconstructed by combining a $K^*$ candidate and a photon candidate. To identify the $B$ mesons, we introduce two kinematic variables: the beam-energy constrained mass, $\Mbc \equiv \sqrt{(E_{\rm beam}^*/c^2)^2 - (p_B^*/c)^2 }$, and the energy difference, $\Delta E \equiv E_B^* - E_{\rm beam}^*$, where $E_{\rm beam}^*$ is the beam energy, and $E_B^*$ and $p_B^*$ are the energy and momentum, respectively, of the $B$ meson candidate in the CM frame. The energy difference is required to be $-0.2$~GeV~$< \Delta E$~$ < 0.1$~GeV; the $\Mbc$ distributions are used to extract the signal yield. 

The dominant background from continuum events is suppressed using a multivariate analysis with a neural network~\cite{NB}. The neural network uses the following input variables calculated in the CM frame: 
(1) the cosine of the angle between the $B$ meson candidate momentum and the $z$ axis, 
(2) the likelihood ratio of modified Fox-Wolfram moments~\cite{SFW,KSFW}, 
(3) the angle between the thrust axes of the daughter particles of the $B$ candidate and all other particles in the rest of the event~(ROE), 
(4) the sphericity and aplanarity~\cite{sph} of particles in the ROE, 
(5) the angle between the first sphericity axes of $B$ candidate and particles in the ROE, 
(6) the absolute value of the cosine of the angle between the first sphericity axes of the particles in the ROE and the $z$ axis, 
and 
(7) the flavor quality parameter of the accompanying $B$ meson that ranges from zero for no flavor information to unity for unambiguous flavor assignment~\cite{TaggingNIM}. 
The output variable, ${\cal{O}}_{\rm NB}$, is required to maximize the significance, defined as $N_S/\sqrt{N_S+N_B}$, where $N_S$ and $N_B$ are the expected signal and background yields for four decay modes in the signal region of 5.27~GeV/$c^2 < \Mbc < $~5.29~GeV/$c^2$, based on MC studies. The criterion ${\cal{O}}_{\rm NB} > 0.13$ suppresses about 89\% of continuum events while keeping about 83\% of signal events for the weighted average of the four decay modes.
The average number of $B$ candidates in an event with at least one candidate is 1.16; we select a single candidate amoung multiple in an event randomly in order not to bias $\Mbc$ and other variables. 
Then, we require the invariant mass of the $K\pi$ system to be within 75~MeV/$c^2$ of the nominal $K^*$ mass. 
The events with invariant mass less than 2.0~GeV/$c^2$ are used to check the contamination from $B \to X_s \gamma$ events that include a higher kaonic resonance decaying to $K\pi$.
The reconstruction efficiencies determined with MC and calibrated by the difference between data and MC with control samples are summarized in Table~\ref{tab:result}.

To determine the signal yields, branching fractions, and direct $CP$ asymmetries in each of the four final states, we perform extended unbinned maximum likelihood fits to the $\Mbc$ distributions within the range $5.20$~GeV/$c^2 < \Mbc < $~5.29~GeV/$c^2$.
The PDF for the signal is modeled by a Gaussian for modes without a $\pi^0$ and a Crystal Ball~(CB) function~\cite{CB} for modes with a $\pi^0$. 
The means of the Gaussian and CB functions are calibrated by $B \to D \pi^-$ events in data while the normalizations and widths are floated. 
The tail parameters of the CB function are determined from signal MC samples. 
From MC studies, it is expected that signal cross-feeds are 0.5\% of the signal yield. We model this cross-feed distribution with a Gaussian and an ARGUS function~\cite{ARGUS}. 
The cross-feed shape and amount of cross-feed relative to correctly-reconstructed signal is fixed to that of the signal MC, such that the cross-feed normalization scales with the signal yield found in data.
The continuum background is described with an ARGUS function. The endpoint of the ARGUS function is calibrated using combinatorial background in $B \to D\pi$ reconstruction in data with the ${\cal{O}}_{\rm NB} < 0.13$ selection to enhance the background statistics; the normalization and the shape parameter are floated. 
The width of the signal and the shape of the ARGUS functions are constrained to be equal between $CP$-conjugate modes but are determined separately across the four subdecay modes.

Backgrounds from $B\bar{B}$ events are small compared with continuum background. However, there are peaking backgrounds mainly from $B \to K\pi\pi \gamma$, $B \to K^*\eta$ and $B^+ \to K^{*+}\pi^0$ events. The $B\bar{B}$ backgrounds are modeled with a bifurcated Gaussian for the peaking component and an ARGUS function for the combinatorial component. The shape and normalization are fixed with large-statistics background MC samples. We take into account the measured $CP$ and isospin violations in the $B\bar{B}$ background~\cite{Olive:2016xmw} to fix the normalizations for $B^+$, $B^-$, $B^0$ and $\bar{B}^0$ mesons.

The likelihood for simultaneous fit over all modes to extract the charged and neutral branching fractions and direct $CP$ asymmetries is defined as
\begin{eqnarray}
& &{\cal{L}}(\Mbc|{\cal{B}}^N, {\cal{B}}^C, A_{CP}^N, A_{CP}^C) \nonumber \\
   &=& \Pi {\cal{L}}^{K_S^0\pi^0}(\Mbc| {\cal{B}}^N) \nonumber \\
   &\times& \Pi {\cal{L}}^{K^-\pi^+}(\Mbc| {\cal{B}}^N, A_{CP}^N) \times \Pi {\cal{L}}^{K^+\pi^-}(\Mbc| {\cal{B}}^N, A_{CP}^N) \nonumber \\
   &\times& \Pi {\cal{L}}^{K^-\pi^0}(\Mbc| {\cal{B}}^C, A_{CP}^C) \times \Pi {\cal{L}}^{K^+\pi^0}(\Mbc| {\cal{B}}^C, A_{CP}^C) \nonumber \\
   &\times& \Pi {\cal{L}}^{K_S^0\pi^-}(\Mbc| {\cal{B}}^C, A_{CP}^C) \times \Pi {\cal{L}}^{K_S^0\pi^+}(\Mbc| {\cal{B}}^C, A_{CP}^C),
\label{eqn:pdf}
\end{eqnarray}
where ${{\cal{L}}^{K\pi}}$ is the likelihood for each final state, and ${\cal{B}}^i$ and $A_{CP}^i$ are the branching fraction and direct $CP$ asymmetry, respectively, in each of the neutral~($N$) and charged~($C$) $B$ mesons. 
Input parameters are the efficiencies for $B^+$, $B^-$, $B^0$ and $\bar{B}^0$ decays, the number of $B\bar{B}$ pairs, $\tau_{B^+}/\tau_{B^0} = 1.076 \pm 0.004$, $f_{+-} = 0.514 \pm 0.006$ and $f_{00} = 0.486 \pm 0.006$~\cite{Olive:2016xmw}. Here, we assume the uncertainties in $f_{+-}$ and $f_{00}$ are perfectly anti-correlated. In the likelihood fit, we can also determine $\Delta A_{CP}$, $\bar{A}_{CP}$ and $\Delta_{0+}$. The combined $A_{CP}(B \to K^{*} \gamma)$ is then obtained by repeating the fit with the constraint $A_{CP}^N = A_{CP}^C$.

The main sources of the systematic uncertainty for the branching fraction measurements are the photon detection efficiency~(2.0\%), the number of $B\bar{B}$ pairs~(1.4\%), the $\pi^0$ detection efficiency~(1.3\%), $f_{+-}/f_{00}$~(1.2\%), and the peaking background yield~(1.1\% to 1.6\%). For the modes with a $\pi^0$ in the final state, fitter bias~($1.3\%$ to $2.4\%$) and fixed parameters in the fit~($1.5\%$ to $3.9\%$)
are also significant sources of uncertainty. The contamination from $B \to X_s \gamma$ events that include a higher-mass kaonic resonance decaying to $K\pi$ is checked by looking at $B \to K \pi \gamma$ events with $M_{K\pi}$ less than 2.0~GeV/$c^2$. The $M_{K\pi}$ distribution is fit with a P-wave relativistic Breit-Wigner for $\Kst$ and a D-wave relativistic Breit-Wigner function for $K_2^*(1430)$ and the resulting uncertainty is 0.31\%. We also check the helicity distribution of the $K\pi$ system for $K^* \gamma$ candidates and find that the distribution is consistent with a P-wave.
For the $\Delta_{0+}$ measurement, the dominant systematic uncertainty is that due to $f_{+-}/f_{00}$~(1.16\%); the second largest is related to particle identification~(0.38\%).
The largest systematic uncertainty for the $A_{CP}$ and $\Delta A_{CP}$ measurements is from the charge asymmetries in charged hadron detection. The charged-pion detection asymmetry is measured using reconstructed $B \to K^{*\pm} \gamma, K^{*\pm} \to K_S^0 \pi^{\pm}$ candidates in ${\cal{O}}_{\rm NB}$ sideband. The charged kaon detection asymmetry is measured using a clean large kaon sample from $D^0 \to K^{+} \pi^{-}$ decay, where the pion detection asymmetry in the decay is subtracted with pions from $D_s^{+} \to \phi \pi^{+}$ decays~\cite{Ko:2012uh}. The raw asymmetries in $B \to K^* \gamma$ are corrected with the measured charged kaon and pion detection asymmetries; $-0.36\pm0.40\%$, $-0.01\pm0.04\%$ and $+0.34\pm0.41\%$ for $K^{+}\pi^{-}$, $K^{+}\pi^0$ and $K_S^0\pi^{+}$ modes, respectively. The second largest is from fitter bias~($0.07\%$ to $0.16\%$) and the third largest is that due to the direct $CP$ asymmetry in rare $B$ meson decays, dominated by $B \to X_s \gamma$, $B \to K^* \eta$ and $B^+ \to K^{*+} \pi^0$~($0.05\%$ to $0.13\%$)~\cite{syst}.

\begin{figure}[htb]
\includegraphics[width=0.238\textwidth]{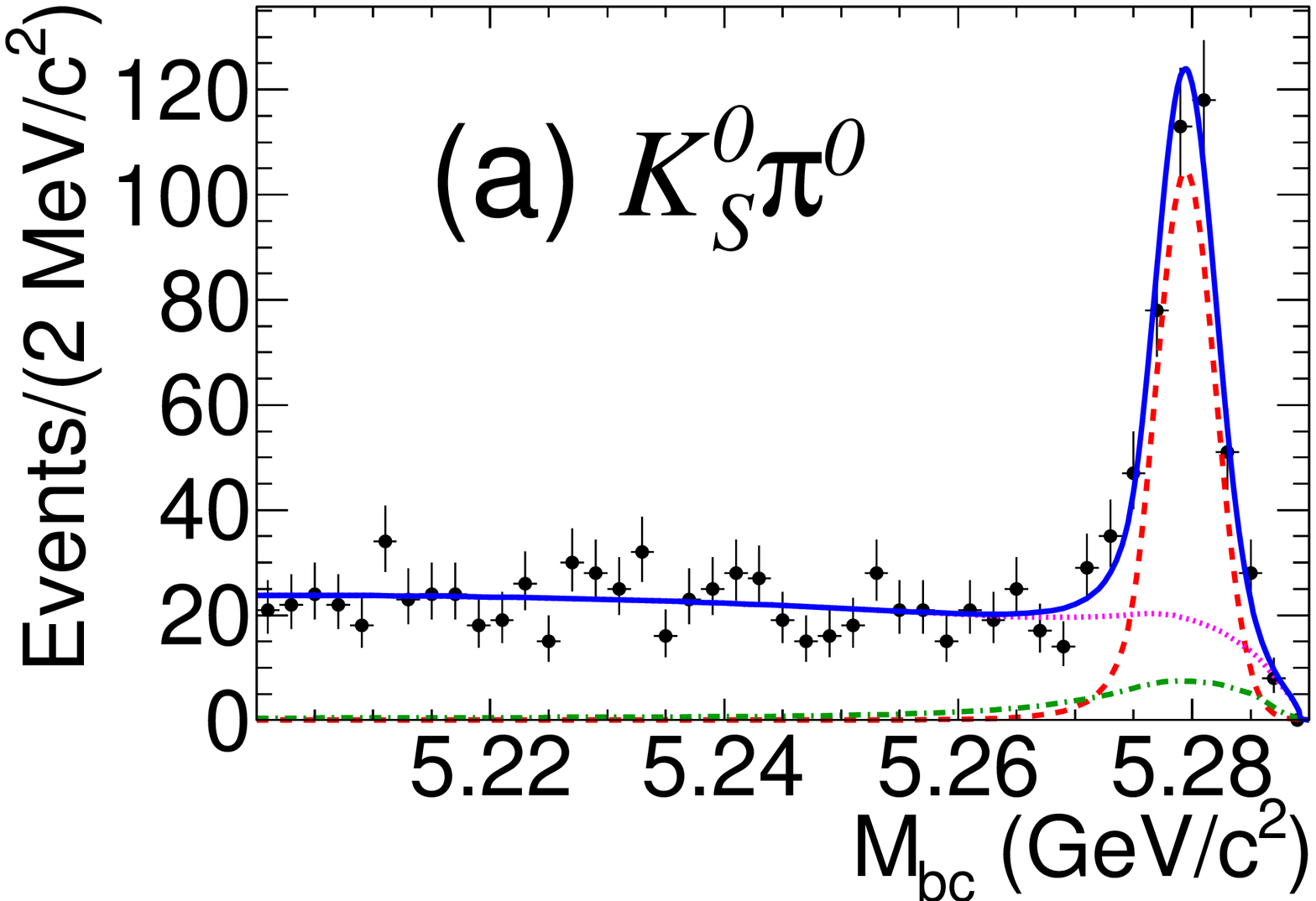}\\
\includegraphics[width=0.238\textwidth]{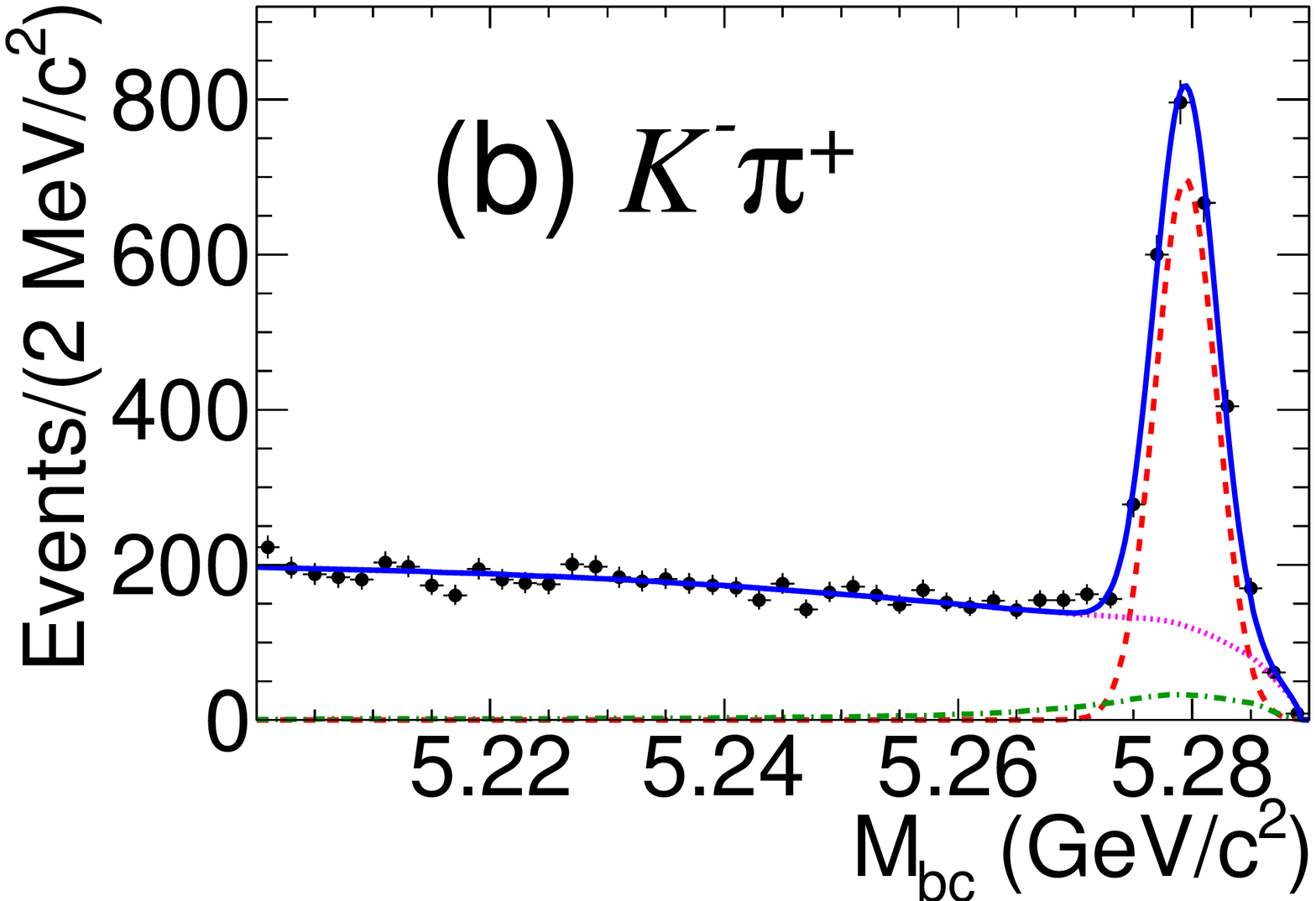}
\includegraphics[width=0.238\textwidth]{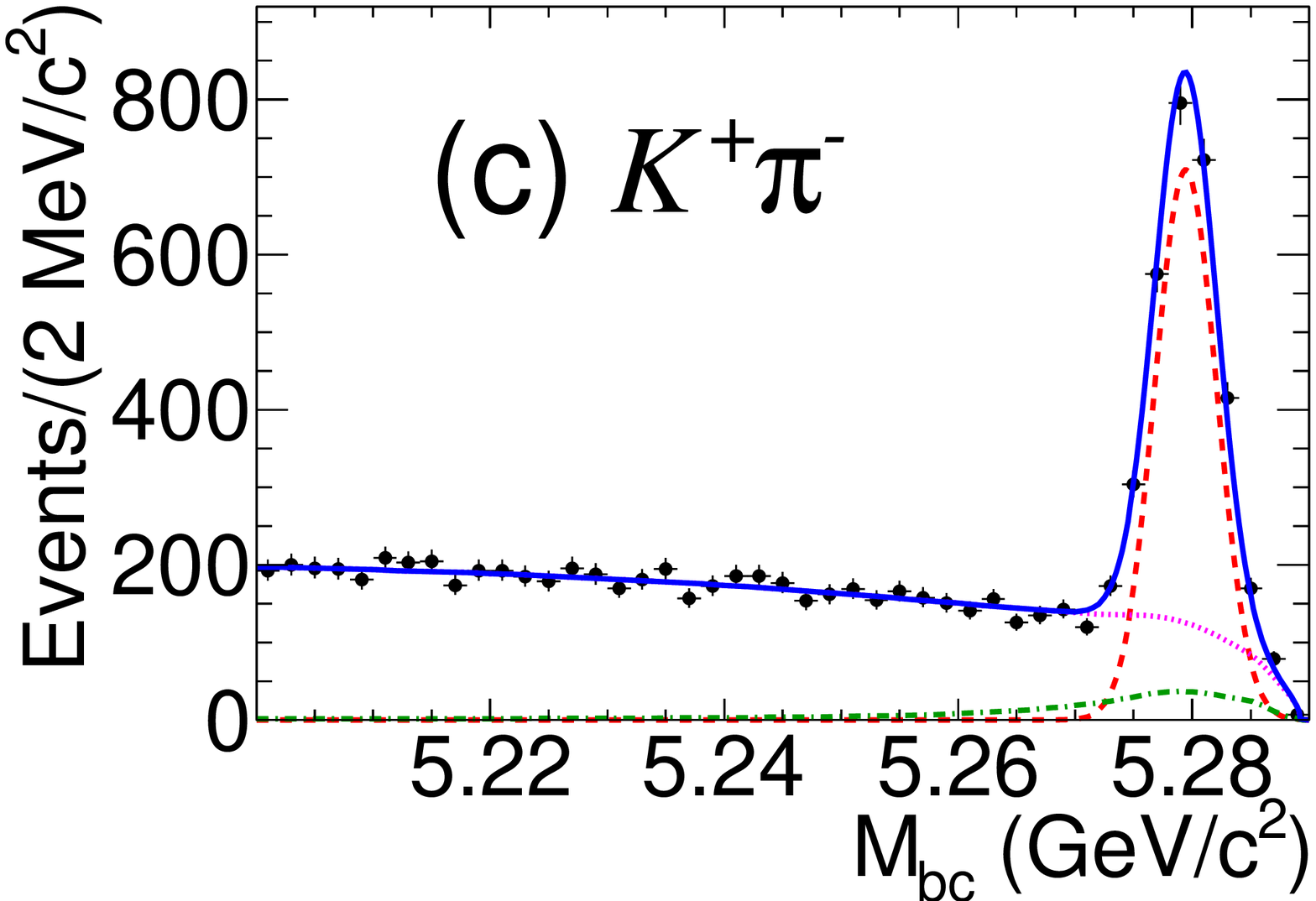}\\
\includegraphics[width=0.238\textwidth]{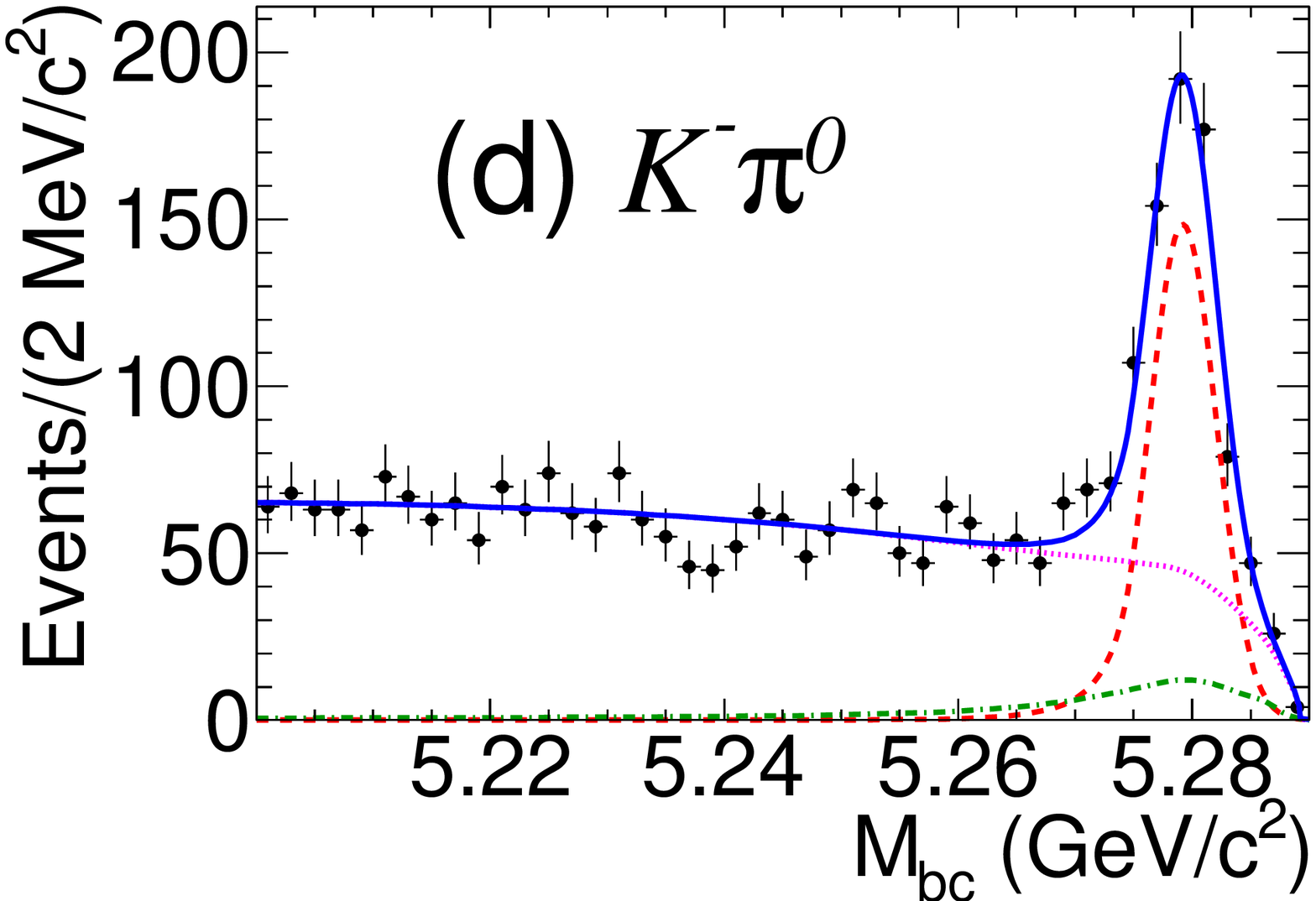}
\includegraphics[width=0.238\textwidth]{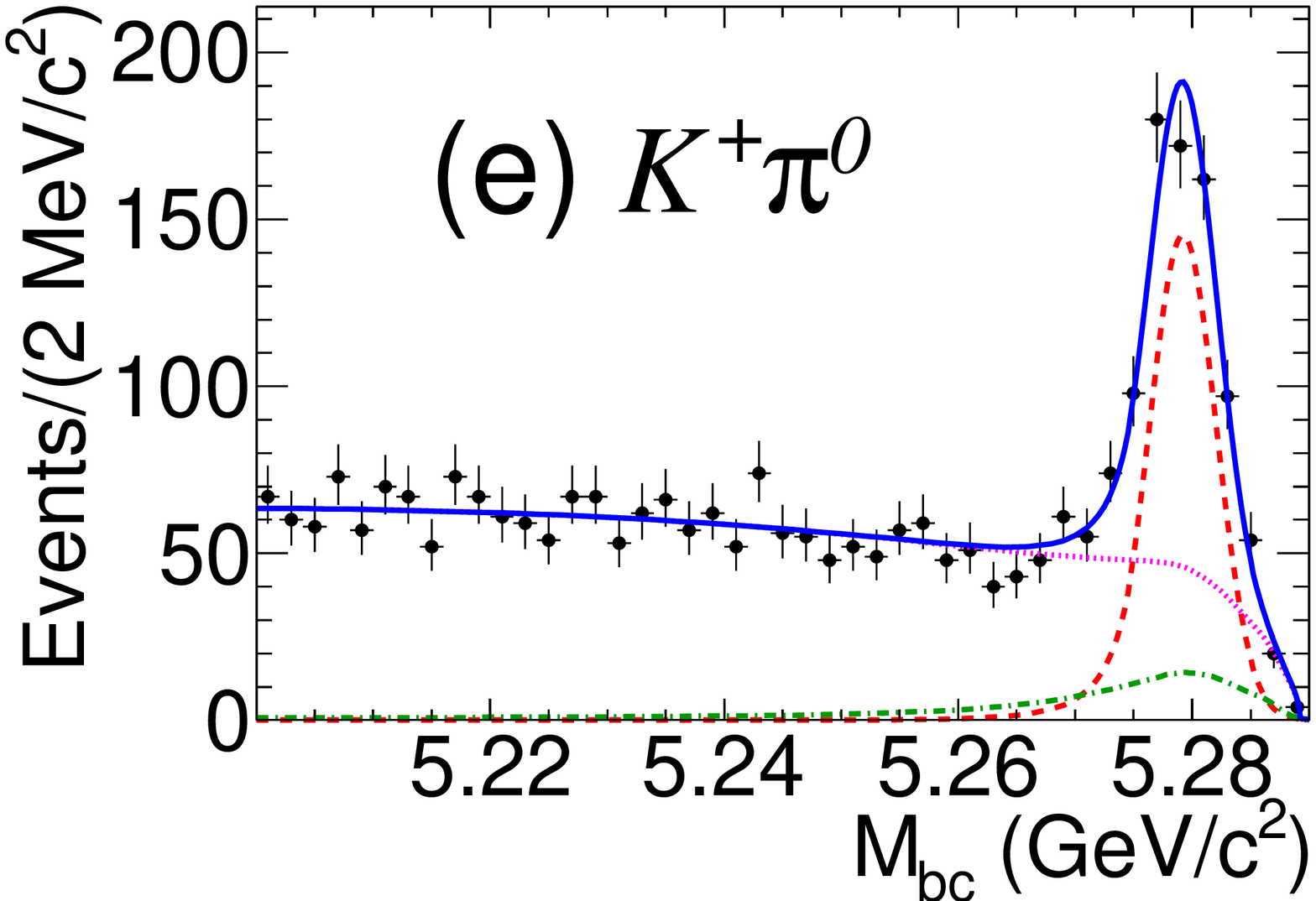}\\
\includegraphics[width=0.238\textwidth]{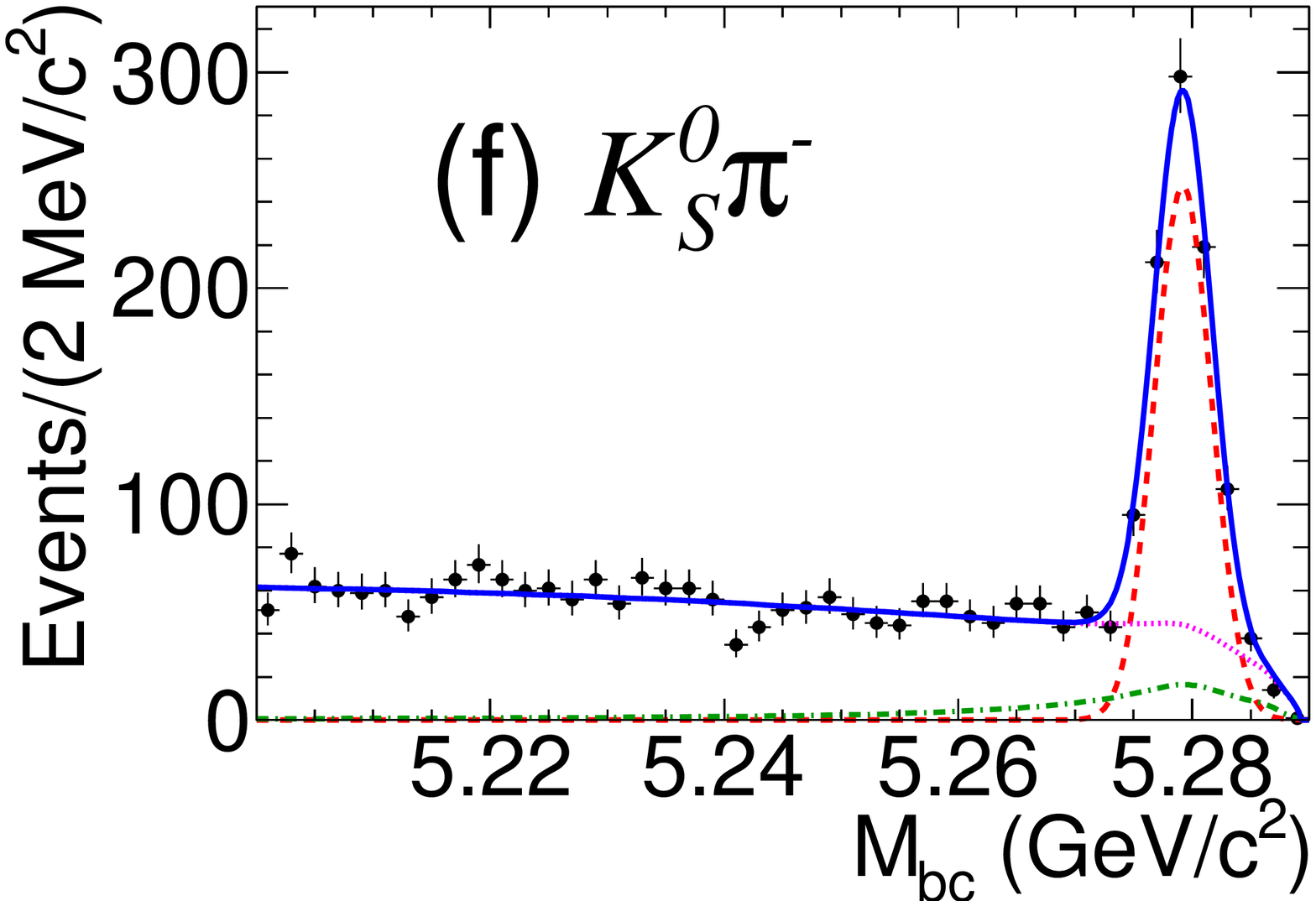}
\includegraphics[width=0.238\textwidth]{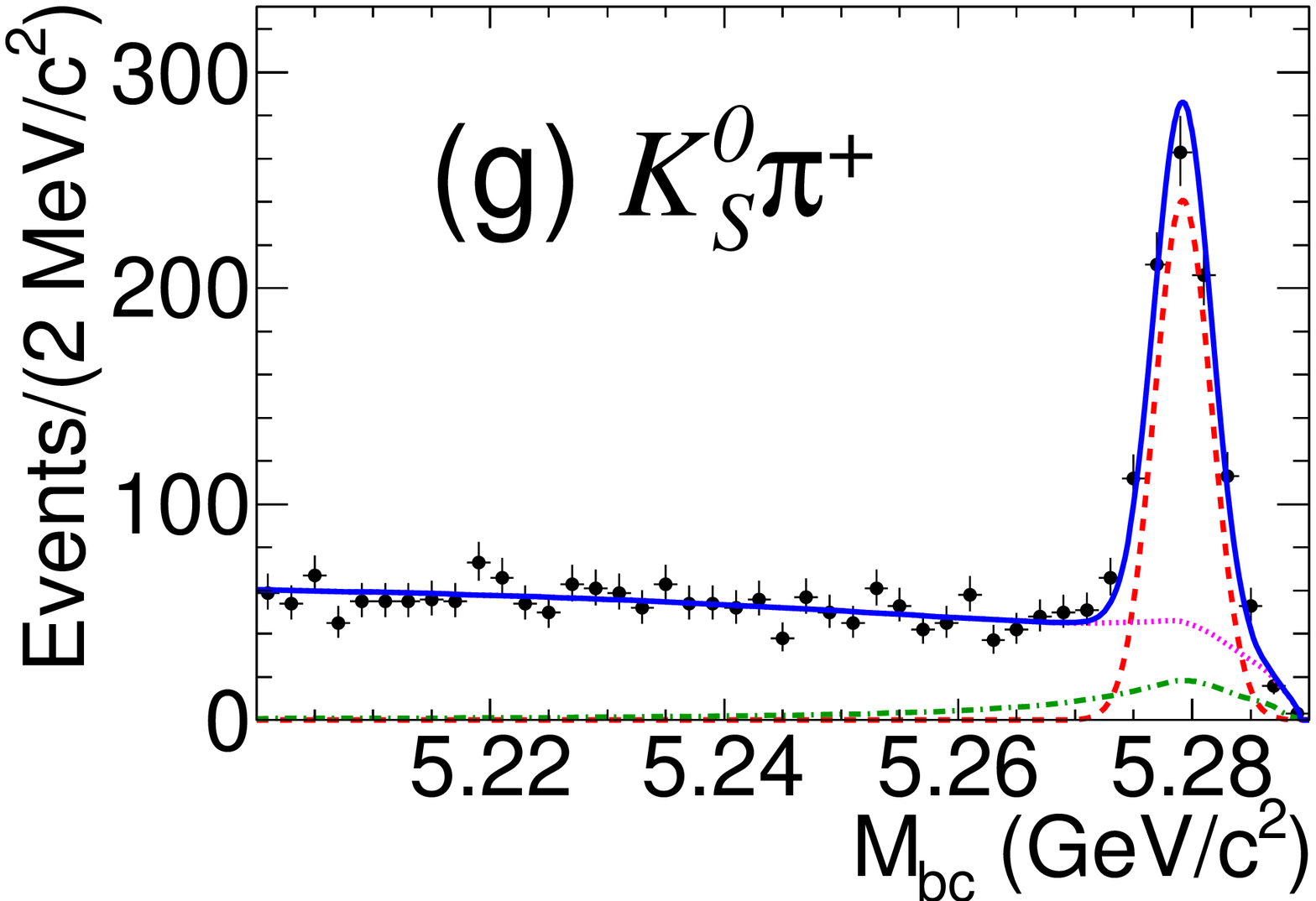}
\caption{$\Mbc$ distributions for (a)~$K_S^0\pi^0$, (b)~$K^-\pi^+$, (c)~$K^+\pi^-$, (d)~$K^-\pi^0$ (e)~$K^+\pi^0$, (f)~$K_S^0\pi^-$ and (g)~$K_S^0\pi^+$. The points with error bars show the data, the dashed (red) curves represent signal, the dotted-dashed (green) curves are $B\bar{B}$ background, the dotted (magenta) curves show total background, and solid (blue) curves are the total.}
\label{fig:mbc}
\end{figure}

First, we extract the branching fraction and direct $CP$ asymmetry in each of the four final states by fitting the $\Mbc$ distributions separated for $\bar{B}$ and $B$ mesons except for the $K_S^0\pi^0$ final state. The results are summarized in Table~\ref{tab:result}. Then, we perform simultaneous fit to seven $\Mbc$ distributions~(Fig.~\ref{fig:mbc}) with the likelihood described above to extract the combined branching fractions and direct $CP$ asymmetries
as well as $\Delta_{0+}$, $\Delta A_{CP}$ and $\bar{A}_{CP}$. The results are
  \begin{eqnarray*}
{\cal{B}}(B^0 \to K^{*0} \gamma) &=& (3.96 \pm 0.07 \pm 0.14)\times10^{-5},\\ 
{\cal{B}}(B^+ \to K^{*+} \gamma) &=& (3.76 \pm 0.10 \pm 0.12)\times10^{-5},\\ 
A_{CP}(B^0 \to K^{*0} \gamma) &=& (-1.3 \pm 1.7 \pm 0.4)\%,\\ 
A_{CP}(B^+ \to K^{*+} \gamma) &=& (+1.1 \pm 2.3 \pm 0.3)\%,\\ 
A_{CP}(B \to K^{*} \gamma)    &=& (-0.4 \pm 1.4 \pm 0.3)\%,\\ 
\Delta_{0+}   &=& (+6.2 \pm 1.5 \pm 0.6 \pm 1.2)\%, \\
\Delta A_{CP} &=& (+2.4 \pm 2.8 \pm 0.5)\%, \\
\bar{A}_{CP} &=& (-0.1 \pm 1.4 \pm 0.3)\%,
  \end{eqnarray*}
where the first uncertainty is statistical, the second is systematic, and the third for $\Delta_{0+}$ is due to the uncertainty in $f_{+-}/f_{00}$~\cite{syst}. 
The $\chi^2$ and number of degrees of freedom in the simultaneous fit calculated from data points and fit curves in Fig.~\ref{fig:mbc} are 256 and 296, respectively.
We find evidence for isospin violation in $B \to K^* \gamma$ decays with a significance of 3.1$\sigma$, and this result is consistent with the predictions in the SM~\cite{Lyon:2013gba,Ball:2006eu,Keum:2004is,Kagan:2001zk,Greub:1994tb,Ahmady:2013cva,Jung:2012vu}. The $A_{CP}$ and $\Delta A_{CP}$ values are consistent with zero. All the measurements are the most precise to date.

\begin{table*}[htb]
\centering
\caption{Signal yields for $\bar{B}$~($N_S^{\bar{B}}$) and $B$~($N_S^B$) mesons, efficiencies ($\epsilon$), branching fractions and direct $CP$ asymmetries. The uncertainties are statistical and systematic except efficiencies. The uncertainties for efficiencies are systematics including statistical uncertainties of MC samples.}
\label{tab:result}
\begin{tabular}{l|ccccc}
\hline \hline
\newlength{\myheight}
\setlength{\myheight}{4mm}
\rule{0cm}{\myheight} Mode               & $N_S^{\bar{B}}$ & $N_S^B$ & $\epsilon$ [\%] & ${\cal{B}}$ [$10^{-5}$] & $A_{CP}$ [\%] \\
\hline
$B^0 \to K_S^0 \pi^0 \gamma$ & \multicolumn{2}{c}{$349 \pm 23 \pm 15$}        & $1.16\pm0.04$  & $4.00 \pm 0.27 \pm 0.24$ & --\\
$B^0 \to K^+ \pi^- \gamma$   & $2295 \pm 56 \pm 27$   &  $2339 \pm 56 \pm 30$ & $15.61\pm0.49$ & $3.95 \pm 0.07 \pm 0.14$ & $-1.3 \pm 1.7 \pm 0.4$ \\
$B^+ \to K^+ \pi^0 \gamma$   & $572 \pm 32 \pm 12$   & $562 \pm 31 \pm 11$    & $3.66\pm0.12$  & $3.91 \pm 0.16 \pm 0.16$ & $+1.0 \pm 3.6 \pm 0.3$ \\
$B^+ \to K_S^0 \pi^+ \gamma$ & $745 \pm 32 \pm 8$   & $721 \pm 32 \pm 9 $    & $5.01\pm0.14$  & $3.69 \pm 0.12 \pm 0.12$  & $+1.3 \pm 2.9 \pm 0.4$ \\
\hline \hline
\end{tabular}
\end{table*}

We also calculate the ratio of branching fractions of $B^0 \to K^{*0} \gamma$ to $B_s^0 \to \phi \gamma$, which is sensitive to annihilation diagrams~\cite{Lyon:2013gba}, based on the branching fraction measurement reported here and the Belle result for the ${\cal{B}}(B_s^0 \to \phi \gamma)$~\cite{Dutta:2014sxo}. To cancel some systematic uncertainties, we take only the $K^+\pi^-$ mode for the branching fractions for $B^0 \to K^{*0} \gamma$. The result is
\begin{eqnarray*}
\frac{{\cal{B}}(B^0 \to K^{*0} \gamma)}{{\cal{B}}(B_s^0 \to \phi \gamma)} &=& 1.10\pm0.16\pm0.09\pm0.18,
\end{eqnarray*}
where the first uncertainty is statistical, the second is systematic, and the third is due to the fraction of $B_s^{(*)0}\bar{B}_s^{(*)0}$ production in $\Upsilon(5S)$ decays.
This result is consistent with predictions in the SM~\cite{Lyon:2013gba,Ali:2007sj} and with LHCb~\cite{Aaij:2012ita}.

%


In summary, we have measured branching fractions, direct $CP$ asymmetries, the isospin asymmetry, and the difference and average of direct $CP$ asymmetries between charged and neutral $B$ mesons in $B \to K^* \gamma$ decays using $772 \times 10^6$ $B\bar{B}$ pairs. We find the first evidence for isospin violation in $B \to K^* \gamma$ with a significance of 3.1$\sigma$. We have made the first measurement of $\Delta A_{CP}$ and $\bar{A}_{CP}$ in $B \to K^* \gamma$ and the result is consistent with zero. The measured branching fractions, direct $CP$, and isospin asymmetries are the most precise to date, and are consistent with SM predictions~\cite{Bosch:2001gv,Lyon:2013gba,Ball:2006eu,Keum:2004is,Kagan:2001zk,Paul:2016urs,Jung:2012vu} and also previous measurements~\cite{Coan:1999kh,Nakao:2004th,Aubert:2009ak,Aaij:2012ita}. These results will be useful for constraining the parameter space in NP models. We also calculate the ratio of $B^0 \to K^{*0} \gamma$ to $B_s^0 \to \phi \gamma$ branching fractions. Current $A_{CP}$ measurements are dominated by the statistical uncertainty; thus, the upcoming Belle~II experiment will further reduce the uncertainty. To observe the isospin violation with 5$\sigma$ significance at Belle~II, reduction of the dominant uncertainty due to $f_{+-}/f_{00}$ is essential, and can be performed at both Belle and Belle~II.

\begin{acknowledgments}
The authors would like to thank Roman Zwicky and David M. Straub for invaluable discussions.
A.~I. is supported by JSPS Grant Number 16H03968 and 
the Munich Institute for Astro- and Particle Physics (MIAPP) of 
the DFG cluster of excellence ``Origin and Structure of the Universe.''
We thank the KEKB group for the excellent operation of the
accelerator; the KEK cryogenics group for the efficient
operation of the solenoid; and the KEK computer group,
the National Institute of Informatics, and the 
PNNL/EMSL computing group for valuable computing
and SINET5 network support.  We acknowledge support from
the Ministry of Education, Culture, Sports, Science, and
Technology (MEXT) of Japan, the Japan Society for the 
Promotion of Science (JSPS), and the Tau-Lepton Physics 
Research Center of Nagoya University; 
the Australian Research Council;
Austrian Science Fund under Grant No.~P 26794-N20;
the National Natural Science Foundation of China under Contracts 
No.~10575109, No.~10775142, No.~10875115, No.~11175187, No.~11475187, 
No.~11521505 and No.~11575017;
the Chinese Academy of Science Center for Excellence in Particle Physics; 
the Ministry of Education, Youth and Sports of the Czech
Republic under Contract No.~LTT17020;
the Carl Zeiss Foundation, the Deutsche Forschungsgemeinschaft, the
Excellence Cluster Universe, and the VolkswagenStiftung;
the Department of Science and Technology of India; 
the Istituto Nazionale di Fisica Nucleare of Italy; 
the WCU program of the Ministry of Education, National Research Foundation (NRF)
of Korea Grants No.~2011-0029457, No.~2012-0008143,
No.~2014R1A2A2A01005286,
No.~2014R1A2A2A01002734, No.~2015R1A2A2A01003280,
No.~2015H1A2A1033649, No.~2016R1D1A1B01010135, No.~2016K1A3A7A09005603, No.~2016K1A3A7A09005604, No.~2016R1D1A1B02012900,
No.~2016K1A3A7A09005606, No.~NRF-2013K1A3A7A06056592;
the Brain Korea 21-Plus program, Radiation Science Research Institute, Foreign Large-size Research Facility Application Supporting project and the Global Science Experimental Data Hub Center of the Korea Institute of Science and Technology Information;
the Polish Ministry of Science and Higher Education and 
the National Science Center;
the Ministry of Education and Science of the Russian Federation and
the Russian Foundation for Basic Research;
the Slovenian Research Agency;
Ikerbasque, Basque Foundation for Science and
MINECO (Juan de la Cierva), Spain;
the Swiss National Science Foundation; 
the Ministry of Education and the Ministry of Science and Technology of Taiwan;
and the U.S.\ Department of Energy and the National Science Foundation.
\end{acknowledgments}

\newpage

The systematic uncertainty for the photon reconstruction efficiency is estimated with radiative Bhabha events. Tracking efficiency uncertainty is estimated using partially reconstructed $D^{*+} \to D^0 \pi^+$, $D^0 \to K_S^0 \pi^+ \pi^-$ events. The uncertainties due to kaon and pion identifications are evaluated with clean kaon and pion samples in $D^{*+} \to D^0 \pi^+$, followed by $D^0 \to K^+ \pi^-$. The uncertainty due to $\pi^0$ reconstruction is determined by taking the ratio of the efficiencies of $\eta \to 3\pi^0$ to $\eta \to \pi^+ \pi^- \pi^0$ or $\eta \to \gamma \gamma$. The uncertainty due to $K_S^0$ reconstruction is evaluated by checking the efficiency of $K_S^0 \to \pi^+ \pi^-$ as functions of flight length, transverse momentum of $K_S^0$ and polar angle of $K_S^0$. The uncertainties due to the ${\cal{O}}_{\rm NB}$ requirement and the $\pi^0/\eta$ veto is estimated with $B \to D \pi$ samples. The uncertainty due to possible mismodeling of the $\Delta E$ distribution is estimated by inflating the $\Delta E$ width and shifting the mean value. The uncertainty due to cross-feed is evaluated by varying the normalization of the PDF by $\pm100\%$ of the nominal value obtained from MC study. The uncertainties due to $A_{CP}$ and $\Delta_{0+}$ in background samples are evaluated by changing these values by $\pm1\sigma$ from the nominal PDG values; if neither $A_{CP}$ nor $\Delta_{0+}$ are not measured, we assign $\pm100\%$ uncertainties. The uncertainty due to the fixed parameters in the fit is evaluated by varying these values by $\pm1\sigma$ of the calibrated values. The fitter bias is checked with a large number of pseudo-MC samples.

\begin{table*}[htb]
\centering
\caption{Systematic uncertainties for branching fractions and $\Delta_{0+}$ in percent. 
For $K\pi$, the results for a separate fit are given while, for $K^*$ and $\Delta_{0+}$, the results for a simultaneous fit are shown. 
}
\label{tab:systbf}
\begin{tabular}{l|ccccccc}
\hline \hline
Source               & $K_S^0\pi^0$ & $K^+\pi^-$ & $K^+\pi^0$ & $K_S^0\pi^+$ & $K^{*0}$ & $K^{*+}$ & $\Delta_{0+}$ \\
\hline
photon reconstruction eff.          &  2.0   &  2.0  &  2.0  &  2.0  &  2.0  &  2.0 & -- \\
tracking eff.                       &  0.7   &  0.7  &  0.4  &  1.1  &  0.7  &  0.8 & 0.05 \\
$K/\pi$ identification eff.         &  --    &  1.7  &  0.8  &  0.8  &  1.6  &  0.8 & 0.38 \\
$\pi^0$ reconstruction eff.         &  1.6   &  --   &  1.6  &  --   &  0.1  &  0.5 & 0.21 \\
$K_S^0$ reconstruction eff.         &  0.2   &  --   &  --   &  0.2  &  $<$0.1  &  0.1 & 0.05 \\
${\cal{O}}_{\rm NB}$ and $\pi^0/\eta$ veto eff.  &  0.6  &  0.6  &  0.6  &  0.6  &  0.6  &  0.6 & -- \\
$\Delta E$ selection eff.          &  1.1  & $<$0.1  &  1.1  &   0.1  &  0.1  &  0.4 & 0.15 \\ 
charge asymmetry in eff.           &  --  &  $<$0.1  &  $<$0.1  &  $<$ 0.1  &  $<$ 0.1  &  $<$0.1 & $<$0.01 \\ 
MC stat.                            &  0.4   &  0.1  &  0.3  &  0.2  &  0.1  &  0.2 & 0.11 \\
number of $B\bar{B}$ pairs          &  1.4   &  1.4  &  1.4  &  1.4  &  1.4  &  1.4 & -- \\
$f_{+-}/f_{00}$                       &  1.2  &  1.2  &  1.2  &   1.2  &  1.2  &  1.2 & 1.16 \\ 
lifetime ratio                      &  --   &  --   &   --  &   --   &  --   &  --  & 0.19 \\ 
higher kaonic resonance             &  0.3  &  0.3  &  0.3  &   0.3  &  0.3  &  0.3 & --   \\
cross-feed                          &  0.2  &  0.2  &  0.3  &   0.2  &  0.2  &  0.2 & 0.03   \\
peaking backgrounds                 &  1.6  &  1.2  &  1.2  &   1.1  &  1.2  &  1.1 & 0.14 \\ 
background $A_{CP}$ and $\Delta_{0+}$ &  0.2  & $<$0.1& $<$0.1&   0.1  &$<$0.1 & $<$0.1 & 0.03 \\ 
fixed parameters in fit             &  3.9  &  0.1  &  1.5  &   $<$0.1  &  0.1  &  0.2 & 0.10 \\ 
fitter bias                         &  2.4  &  0.2  &  1.3  &   0.7  &  0.2  &  0.2 & 0.08 \\ 
\hline
total                               &  5.9  &  3.5  &  4.2  &   3.3  &  3.5  &  3.3 & 1.29 \\ 
\hline \hline
\end{tabular}
\end{table*}

\begin{table*}[htb]
\centering
\caption{Systematic uncertainties for $A_{CP}$, $\Delta A_{CP}$ and $\bar{A}_{CP}$ in percent. 
For $K\pi$, the results for a separate fit are given while, for $K^*$ and $\Delta A_{CP}$, the results for a simultaneous fit are shown.
Systematic uncertainies due to tracking, $K/\pi$ identification, and $\pi^0$ and $K_S^0$ reconstruction efficiencies are only accounted
for in the simultaneous fit results since the uncertainties of the relative efficiencies of the decay modes change the fit results.
}
\label{tab:systa}
\begin{tabular}{l|cccccccc}
\hline \hline
\setlength{\myheight}{4mm}
\rule{0cm}{\myheight} Source                                     & $K^+\pi^-$ & $K^+\pi^0$ & $K_S^0\pi^+$ & $K^{*0}$ & $K^{*+}$ & $K^*$ & $\Delta A_{CP}$ & $\bar{A}_{CP}$ \\
\hline
tracking eff.                             &   --   &   --  &  --  &   $<$0.01 &  $<$0.01 &  $<$0.01 & $<$0.01 & $<$0.01  \\
$K/\pi$ identification eff.               &  --    &   --  &  --  &   $<$0.01 &  $<$0.01 & $<$0.01 & $<$0.01 & $<$0.01  \\
$\pi^0$ reconstruction eff.               &  --    &  --   &  --  &  $<$0.01 & $<$0.01 &  $<$0.01  &  $<$0.01 &  $<$0.01  \\
$K_S^0$ reconstruction eff.               &  --   &  --   &  --   &  $<$0.01 & $<$0.01  &  $<$0.01 & $<$0.01 & $<$0.01  \\
charge asymmetry in $K/\pi$ detection     &  0.40 &  0.04  &  0.41 &  0.40 &  0.25 &  0.28 & 0.48  & 0.24 \\ 
cross-feed                                &  0.02 &  0.04  &  0.03 &  0.02 &  0.02 &  0.02 & 0.02  & 0.01 \\ 
peaking backgrounds                       &  0.04 &  0.06  &  0.08 &  0.04 &  0.06 &  0.05 & 0.04  & 0.05 \\ 
background $A_{CP}$ and $\Delta_{0+}$       &  0.10 &  0.13  &  0.09 &  0.10 &  0.10 &  0.10 & 0.05  & 0.10\\ 
fixed parameters in fit                   &  $<$0.01 &  0.13  &  0.02 &  $<$0.01 &  0.02 &  $<$0.01 & 0.02 & 0.06 \\ 
fitter bias                               &  0.07 &  0.16  &  0.12 &  0.07 &  0.09 &  0.08 & 0.12 & 0.06 \\ 
\hline
total                                     &  0.42 &  0.26  &  0.45 &  0.42 &  0.30 &  0.31 & 0.50 & 0.27\\ 
\hline \hline
\end{tabular}
\end{table*}

\newpage
The correlation matrix including statistical and systematic effects for seven observables is shown.

\begin{table*}[htb]
\centering
\caption{The correlation matrix for seven observables. The ${\cal{B}}^i$ and $A_{CP}^i$ are the branching fraction and direct $CP$ asymmetry, respectively, in each of the neutral~($N$) and charged~($C$) $B$ mesons.
}
\label{tab:cor}
\begin{tabular}{l|ccccccc}
\hline \hline
\setlength{\myheight}{4mm}
\rule{0cm}{\myheight} & ${\cal{B}}^N$ & ${\cal{B}}^C$ & $A_{CP}^N$ & $A_{CP}^C$ & $\Delta_{0+}$ & $\Delta A_{CP}$ & $\bar{A}_{CP}$ \\
\hline
${\cal{B}}^N$      &  1.00  &  0.49  &  -0.01  &   0.00  &  0.46  &   0.00  &  -0.01  \\
${\cal{B}}^C$      &  0.49  &  1.00  &  -0.01  &  -0.01  & -0.55  &   0.00  &  -0.01  \\
$A_{CP}^N$          & -0.01  & -0.01  &   1.00  &   0.00  &  0.00  &  -0.56  &   0.59  \\
$A_{CP}^C$          &  0.00  & -0.01  &   0.00  &   1.00  &  0.01  &   0.76  &   0.80  \\
$\Delta_{0+}$       &  0.46  & -0.55  &   0.00  &   0.01  &  1.00  &   0.01  &   0.00  \\
$\Delta A_{CP}$     &  0.00  &  0.00  &  -0.56  &   0.76  &  0.01  &   1.00  &   0.27  \\
$\bar{A}_{CP}$      & -0.01  & -0.01  &   0.59  &   0.80  &  0.00  &   0.27  &   1.00  \\
\hline \hline
\end{tabular}
\end{table*}


\begin{thebibliography}{99}
\bibitem{Kstar}
  The $\Kst$ is denoted as $K^*$ throughout this Letter.
\bibitem{CC}
Throughout this paper, the inclusion of the charge conjugate mode decay is implied unless otherwise stated.
\end{thebibliography}
\end{document}